\documentstyle[12pt]{article}

\newcommand{\beq}{\begin{equation}}
\newcommand{\eeq}{\end{equation}}
\newcommand{\be}{\begin{equation}}
\newcommand{\ee}{\end{equation}}
\newcommand{\beaa}{\begin{eqnarray}}
\newcommand{\eeaa}{\end{eqnarray}}
\newcommand{\bea}{\begin{eqnarray}}
\newcommand{\eea}{\end{eqnarray}}
\newcommand{\ba}{\begin{array}}
\newcommand{\ea}{\end{array}}

\def\le{\left}
\def\ri{\right}

\def\lrar{\leftrightarrow}

\def\rar{\rightarrow}

\def\lab{\label}

\def\lan{\langle}
\def\ran{\rangle}

\def\bt{\beta}

\def\om{\omega}

\def\si{\sigma}

\newcommand{\half}{\frac{1}{2}}

\newcommand{\veck}{{\bf {k}}}
\newcommand{\vecp}{{\bf {p}}}
\newcommand{\vecx}{{\bf {x}}}

\newcommand{\fnorm}{\frac{1}{\sqrt{2 \omega_k V}}}

\newcommand{\A}{{a}}

\newcommand{\ak}{a_{\veck}}
\newcommand{\adk}{a^{\dagger}_{\veck}}
\newcommand{\amk}{a_{-\veck}}
\newcommand{\admk}{a^{\dagger}_{-\veck}}

\newcommand{\Wp}{{W}_1}
\newcommand{\Wm}{{W}_2}
\newcommand{\WFp}{{W}^\flat_1}
\newcommand{\WFm}{{W}^\flat_2}

\newcommand{\href}[2]{{#2}{}}
\newcommand{\eprint}[1]{\href{http://xxx.soton.ac.uk/abs/#1}{({\tt #1})}}
\newcommand{\eprintpub}[1]{\eprint{#1}}

\newcommand{\tseref}[1]{(\ref{#1})}
\newcommand{\tsenote}[1]{ }

\begin{document}

\thispagestyle{empty}
\begin{flushright} 
\href{http://euclid.tp.ph.ic.ac.uk/Papers/papers_96-7_.html}{Imperial/TP/96-97/69} \\ 
\eprint{hep-ph/9706549} \\ 
12th January 1998\\
(\LaTeX-ed on \today )
\end{flushright}

\vspace{2.0cm}

\centerline{\large{\bf On Normal Ordering and Canonical 
Transformations} }

\vspace{0.2cm}

\centerline{\large{\bf in Thermal Field Theory}\tsenote{All notes of
this type are not for public viewing, and can be removed by
redefining the {\tt tsenote} command in the preamble.} } 

\vspace{1.5cm}

\begin{center}
{\large M. Blasone$^{a}$\footnote{E-mail: {\tt blasone@pcvico.csied.unisa.it}} , 
T.S. Evans$^{b}$\footnote{E-mail: {\tt T.Evans@ic.ac.uk},
WWW: \href{http://euclid.tp.ph.ic.ac.uk/timeindex.html}{{\tt http://euclid.tp.ph.ic.ac.uk/$\sim$time}}},
D.A. Steer$^{b}$\footnote{E-mail: {\tt D.Steer@ic.ac.uk}} 
and 
G. Vitiello$^{a}$\footnote{E-mail: {\tt vitiello@vaxsa.csied.unisa.it}} } \\[0.5cm]
{\it a) Dipartimento di Fisica, Universit\`a di Salerno, 84100 
Salerno, 
Italy,\\
INFN Gruppo Collegato di Salerno}\\
{\it b) Blackett Laboratory, Prince Consort Road, Imperial College,
London, SW7 2BZ, U.K.}
\end{center}

% **************************************************************
%\typeout{Details for Publisher included}
%\vspace{1.5cm}
%\begin{center}
%Corresponding Author: T.S. Evans$^{b2}$ \\
%Tel: +44-171-594-7837 \\
%Fax: +44-171-594-7844 \\
%\mbox{  } \\
%PACS: 11.10.Wx, 11.10-z \\
%\mbox{ } \\
%{\bf Keywords} \\
%Bogoliubov Transformations, Finite Temperature, \\ 
%Canonical Transformations, Normal Ordering
%\end{center}
% **************************************************************

\vspace{1.5cm}

\centerline{{\bf Abstract}}  
We look at a real scalar field in thermal equilibrium in the context
of the new normal ordering and field split defined by Evans and
Steer \cite{ES}.  We show that the field split defines a natural
canonical transformation, but that this transformation differs from
others known in thermal field theory.

%\newpage
%\setcounter{page}{1}
%\setcounter{equation}{0}

\section{Introduction}

In some recent work \cite{ES}, the concept of normal ordering in
path ordered approaches to thermal field theory
\cite{Ma,Raybook,LvW,vW,LeB}  was discussed.  In particular it was
shown that with a new definition, one can ensure that the thermal
expectation value of all normal ordered products are zero.  This is
true for all types of fields and for all contours in the complex
time plane, including Matsubara's imaginary time contour \cite{Ma}. As a
result, the canonical derivation of quantum field theory (QFT) at
finite temperature, $T$, can be seen to proceed just as at zero
temperature. In \cite{ES} normal ordering was defined in terms of
the split of the field, say ${\psi}(x)$, into ``positive'' and
``negative'' parts:  
\be
\psi(x) = \psi^{(+)}(x) + \psi^{(-)}(x).
\lab{decomp}
\ee
In the following discussion we keep the labels ``positive'' and
``negative'' even though, as was shown in \cite{ES}, the preferred
$T>0$ split is more general than the usual $T=0$ split and is {\em not} a
split into positive and negative energy waves.

We remark that the use of a more general split is natural since in QFT
the main task of field splitting is not so much the separation of
the field into positive and negative energy waves, but its
separation into annihilation and creation operators.  Thus the real
problem is which {\em definition} of the annihilation and creation
operators to use in order to best describe the physical system under
consideration.   In other words, it is important to choose the
correct physical representation of the underlying algebraic
structure of the theory, since the representations have a different
physical content and the algebra alone does not determine the
annihilation and creation operators.  The fundamental property of
QFT is that there are infinitely many such representations, and this
allows it to represent a wide variety of physical phenomena
\cite{St}.

There are many examples of this.  In spontaneous  symmetry breaking,
the  value of the order parameter {\em selects} the appropriate
physical {\it  vacuum} and thus the appropriate representation of
the canonical  commutation relations.  This means that the
irreducible set of  physical fields is not given {\it a  priori},
but is dictated by the symmetry breaking condition \cite{St}.  
In a similar way, in flavour oscillations (such as in kaon or
neutrino systems) a non-trivial choice of appropriate physical
vacuum is made \cite{mix}.  This problem must also be faced
when working with quantum fields in curved space-time \cite{MSV,BD}
or when studying the canonical quantisation of gravitational wave
mode evolution in   inflating universe \cite{AMV}.  Closely
related is the need to pick representations carefully when
quantising dissipative systems \cite{QD}.  In all the above
cases, Bogoliubov transformations \cite{BV}  play a central role,
and the new creation and annihilation operators defined by such
transformations are invariably related by hermitian
conjugation.\footnote{In some of the above situations, since the
Lorentz (Poincar\'e) covariance is   lost, the notion of vacuum (and
thus of field splitting into positive and   negative frequency
parts) is in fact missing. And this brings us back to the problem 
of the {\it definition} of the annihilation and creation 
operators\cite{MSV}
and of the  {\it proper} normal ordering.}  Finally, we note that
careful attention is needed to define the {\it proper} creation 
and annihilation operators in the quantisation of two-dimensional  
gravity models \cite{ja}.

Returning to the thermal field theory context, it is well known  
that the very same particle concept loses any meaning at non-zero  
temperature \cite{La,He}. There is, therefore, an intrinsic physical
relevance in the   study of how to define and to deal with {\em
particle} creation and   annihilation operators, if any, at finite
temperature, and in the   understanding of which extent it is
possible to associate a meaning to   them in terms of physical
excitations. In a word, the problem of how to   construct a
formalism which is `canonical' is a crucial, and in some respects an
  urgent problem to solve. The discussion we present in this paper
is a   contribution in such a direction.  Our stand-point is that,
in a   similar way to Classical Mechanics and to zero temperature
QFT,   searching for and studying the properties of the canonical  
transformations of the theory is priority in order to keep contact
with   the physical content of the formalism.

It is well known that in thermal field theory,
Bogoliubov transformations are central to the approach known as
TFD (Thermo Field Dynamics) \cite{Raybook,LvW,vW,He,TFD}.  There the
correct choice of vacuum depends on the temperature, but
otherwise the derivation is very similar to the canonical approach
to zero temperature QFT.  An interesting point to
note -- which will become relevant later -- 
is that in TFD it is possible to work with a pair of canonical
operators which are {\em not} hermitian conjugates ($\alpha \neq
1/2$ in the notation of \cite{He}).  Such non-hermitian
representations give the same physical results at equilibrium and
indeed one ($\alpha=1$) is favoured both in and out of equilibrium
\cite{vW,He}.  

Given the prevalence of Bogoliubov transformations in quantum field
theory and their central role in the TFD approach to thermal field
theory, it is surprising that no such structure has been found
in the alternative path-ordered approaches to thermal field theory,
which is the context of the new normal ordered product of
\cite{ES}. The path-ordered formalisms are all based on a contour
in the complex time-plane which ends $-i \beta$ below its starting
point \cite{Ma,Raybook,LvW,vW,LeB}.  They come in two varieties,
real- and imaginary-time formalisms.  The real-time versions are
distinct from TFD \cite{LvW} (for example the various fields and their
creation and annihilation operators have distinct properties) 
but perturbatively they are
completely equivalent in equilibrium, thus reflecting an underlying relationship
\cite{vW,Louis}.\footnote{This
similarity has led to confusion in the literature over
nomenclature.}  In the real-time
path-ordered approaches a thermal Bogoliubov transformation
appears, but only post-hoc, say when the structure of the propagator is
analysed.  However, there is no sign of a thermal Bogoliubov
transformation in the imaginary-time approaches.

The split discovered in \cite{ES} by considering normal ordering
in path ordered TFT combines annihilation and creation operators in
a way that is reminiscent of Bogoliubov transformations used in
other problems.  The aim of this paper is to elucidate the precise
relation between the split of \cite{ES} and canonical
transformations.  In section 2, we introduce our notation and  
summarise the results of \cite{ES} focusing on real scalar
fields.  In section 3 we search for a canonical structure in those
results.   This structure is then put in the context of Bogoliubov
transformations in section 4.  In section 5 we present our
conclusions, and discuss extensions of our work to other fields.

\section{Normal ordering and Field Splitting }

We consider real scalar relativistic fields
\be\lab{phi}
\phi(x) =
\sum_{ \bf k} \frac {1}{\sqrt{ 2 \om_{\bf k} V}}
\le(a_{{\bf k}}(t)\,e^{i {\bf k\cdot x}} \, + 
\, a^{\dagger}_{{\bf k}}(t)e^{-i {\bf k\cdot x}}\ri) ,
\ee
where $a_{{\bf k}}(t) =e^{-i \om_{\bf k} t} a_{{\bf k}}$ and $V$ is
the volume of the system.
The precise form of the dispersion relation is unimportant; we only 
require that $\om_{\bf k} = \om_{-{\bf k}}$. 
%However, to simplify the notation we work with the simpler case 
%$\om_{\bf k} = \om_{|\veck|}$.  
The annihilation and creation operators ${a}_{\bf
k}$, ${a}^{\dagger}_{\bf k}$ obey the canonical commutation relations 
\be\lab{comma}
\le[ {a}_{\bf k}, {a}^{\dagger}_{\bf k'} \ri]  = 
\delta_{{\bf k},{\bf k'}} ,
\ee
and all other commutators are zero.  The vacuum 
state for $ {a}_{\bf k}$ is denoted by $|0\rangle$; $ {a}_{\bf
k}|0\rangle = 0$.  We consider thermal equilibrium and denote
thermal averages by double angular brackets:
\be\lab{3bis}
\langle \langle \ldots \ran \ran   =  \frac{1}{Z} {\rm Tr} \le\{ e^{-\bt H} 
\ldots \ri\}.
\ee
Here $\ldots$ means any operator, ${H}$ is the 
Hamiltonian, $Z$ is the
partition function and $\bt$ is the inverse temperature: $\bt
= 1/T$ ($k_B = 1$).  The trace is over a complete set of states for the
system.  Since we have no chemical potential
\bea
\langle \langle {a}^{\dagger}_{\bf k} {a}_{\bf p} \ran \ran  =
n_{\bf k} \delta_{{\bf k},{\bf p}} \; , &&
\langle \langle {a}_{\bf k} {a}^{\dagger}_{\bf p} \ran \ran  =  
\le( 1 +  n_{\bf k} \ri) \delta_{{\bf k},{\bf p}}\, ,
\lab{b1}
\eea 
where $n_{\bf k}$ is given by the Bose-Einstein distribution
\bea 
n_{\bf k} = {1\over { e^{\bt \om_{\bf k} } - 1  }}
\lab{BEFD}.
\eea

In \cite{ES} normal ordering was defined in terms of the arbitrary 
split \tseref{decomp}, so generalising the traditional $T=0$ 
definition.  
We {\em always} define normal ordering to strictly mean that all
$(+)$-fields are moved to the right of $(-)$-fields; otherwise the 
order of the fields is left unchanged. 
For example,
\be 
N \le[ \phi_1 \phi_2 \ri]  =
\phi_{1}^{(+)} \phi_{2}^{(+)} +
\phi_{1}^{(-)} \phi_{2}^{(+)} +   \phi_{2}^{(-)} \phi_{1}^{(+)} +
\phi_{1}^{(-)} \phi_{2}^{(-)},
\lab{N2}
\ee 
where $\phi_i = \phi(\vecx_i,t_i)$.  

Using this generalised definition of normal ordering, it was shown
in  \cite{ES} that Wick's theorem holds in its usual form {\em
if} a split is chosen such that the contraction is a $c$-number. 
This is satisfied by  splits \tseref{decomp} which are linear in the
annihilation and creation operators and only these were considered
in \cite{ES}.  It was then shown that if the fields  are
split such that the thermal expectation value of two-point normal
ordered  products vanish, then the thermal expectation value of all
$n$-point  normal ordered products vanish. 

The only splits which guarantee that 
$\langle \langle  N \le[ {\phi}_1 {\phi}_2 \ri] \ran \ran = 0$ 
for all times were shown in \cite{ES} to be
\begin{eqnarray}
{\phi}^{(+)}(x) &=& \sum_{\veck} \fnorm
\left[ (1 - f_{\veck}) {a}_{\veck} e^{-ik.x} + g_{\veck}
{a}^{\dagger}_{\veck} e^{ik.x} \right],
\label{phiplus}
\\
{\phi}^{(-)}(x) &=& \sum_{\veck} \fnorm
\left[ f_{\veck} {a}_{\veck} e^{-ik.x} + (1 - g_{\veck})
{a}^{\dagger}_{\veck} e^{ik.x} \right],
\label{phiminus}
\end{eqnarray}
where there are two solutions for $f_{\bf k}$ and $g_{\bf k}$, namely
\bea
f_{\bf k} = -  n_{\bf k} + s_{\bf k}
[n_{\bf k}(n_{\bf k}+ 1 )]^{1/2} ,
&&
g_{\bf k} = -  n_{\bf k} - s_{\bf k} [n_{\bf k}(n_{\bf k} + 1)]^{1/2},
\lab{fgdef1}\\
s_{\bf k} &=& \pm 1  .
\eea
Note that $s_{\bf k}$ can be chosen to be a
function of
both the size {\em and} direction of ${\bf k}$.   

These solutions were derived within the context of the path-ordered
approaches to thermal field theory.  The solution is independent of
the curve chosen and so all the work in \cite{ES} and in the
present paper applies to both imaginary-time and all real-time
path-ordered approaches to thermal field theory.

\section{Searching for Canonical Transformations}

The split \tseref{phiplus}-\tseref{phiminus} contains factors which
are reminiscent of those seen in Bogoliubov transformations, and
particularly those encountered in TFD.  It is therefore interesting to
see if this new split defines any new canonical operators and if so,
whether those operators can be given any physical significance.

We may rewrite the split \tseref{phiplus}-\tseref{phiminus} as
\be
\phi^{(+)} = \sum_{\bf k} 
\frac{d_{\bf k}}{\sqrt{2 \om_{\bf k} V}}
 e^{+i {\bf k}.{\bf x}} X_{\bf k}(t) ,
\; \; \;
\phi^{(-)} = \sum_{\bf k} 
\frac{d_{\bf k}}{\sqrt{2 \om_{\bf k} V}}
e^{-i {\bf k}.{\bf x}} X_{\bf k}^{\flat}(t) .
\lab{ppr}
\ee
where $d_{\bf k}$ is a normalisation factor to be determined later,
and the new operators $X_{\bf k}$ and $X^{\flat}_{\bf k}$ are given by
\bea
X_{\bf k}(t) &=& e^{- i \om_{\bf k} t}\; \frac{(1-f_{\bf k})}{d_{\bf k}
}
\; a_{\bf k}  \;  + \;
e^{+ i \om_{\bf k} t}\; \frac{g_{\bf k}}{ d_{\bf k}}\;  a_{\bf -
k}^{\dagger} ,
\\ 
X^\flat_{\bf k}(t) &=& e^{- i \om_{\bf k} t}\; \frac{f_{\bf k}}{d_{\bf
k} }
\; a_{\bf - k}  \; +\; 
e^{+ i \om_{\bf k} t}  \; 
\frac{(1-g_{\bf k})}{d_{\bf k}} \; a_{\bf k}^{\dagger}  .
\lab{X}
\eea
Note that the new definition of normal ordering, \tseref{N2},
\tseref{phiplus} and \tseref{phiminus},  is now
equivalent to the rule of putting all `annihilation' (`creation')
operators $X_{\bf k}$ ($X_{\bf k}^\flat$) to the left (right), 
just as we are used
to at zero temperature.  We know from the work of \cite{ES} that
this will guarantee  that the thermal expectation value of any
normal ordered product of fields vanishes.  Thus for this finite
temperature system,  $X_{\bf k}$ and $X_{\bf k}^{\flat}$ seem to
mimic the usual creation and annihilation operators. 

We have introduced a new operation, which we call
`flat conjugation' and denote with a $\flat$
symbol.\footnote{Throughout this paper we use musical symbols
to denote non-hermitian operations. This follows Henning
\cite{He} who uses the `sharp' symbol, $\sharp$, 
to cope with the non-hermitian nature
of $\alpha \neq 1/2$ TFD representations.} Flat conjugation is
defined by \tseref{X} and it consists of both  hermitian conjugation
and the  exchange $f \lrar g$: 
\be
A^{\flat}[f,g]  = \left( A[f,g] \right)^{\flat} \equiv 
 \left( A[g,f] \right)^{\dagger} 
\lab{flatdef}
\ee
for any operator ${A}$, so that $\le( {A}^{\flat}
\ri)^ {\flat} = {A}$.  The $\flat$-operation is needed because it, and
not hermitian conjugation, now relates the
two parts of the field,
\be
\le[\phi^{(+)}\ri] ^{\flat} = \phi^{(-)}  \neq
\le[\phi^{(+)}\ri] ^{\dagger}.
\lab{fTflat} 
\ee
Only at $T=0$, where $f_{\bf k}=g_{\bf k}=0$, does $\flat = \dagger$ from
\tseref{flatdef}, and there we recover the usual relationship
\be
\le[\phi^{(+)}\ri] ^{\flat} = \le[\phi^{(+)}\ri]
^{\dagger} = \phi^{(-)}
\lab{T0flat}.
\ee

If we also enforce that the $X_\veck$ operators satisfy 
equal time canonical commutation relations
\be
[X_{\bf k}(t), X_{\bf p}^\flat (t)] = \delta _{{\bf k},{\bf p}}
\lab{x1},
\ee
\be
[X_{\bf k}(t) , X_{\bf p}(t) ]  = 0 = [X_{\bf k}^{\flat}(t) , X_{\bf
p}^{\flat}(t) ] .
\lab{x2}
\ee
We find that the last two equations may only be satisfied if we
choose $s_{\bf k} = s_{-\bf k}$.  This determines 
the normalisation factor, $d_{\bf k}$, to be 
\be
d_{\bf k} =d_{-{\bf k}}= \left(1 + 2n_{\bf k}\right)^{1/2}.
\lab{ddef}
\ee
%
%which also shows that $d_{\bf k}$ depends only on the $\bf k$ modulus. 
Thus the normalisation factor which appears in the fields \tseref{ppr}
is the square root of
\be
\frac{d_{\bf k}^2}{{2 \om_{\bf k} V}}
\; = \; \frac{1}
{2 \om_{\bf k} V} \coth \le( \frac{\bt \om_{\bf k}}{2}
\ri) = \frac{1}{V} \int dk_0  [ \theta(k_0) + n_{|k_0|} ]
\delta(k_0^2-\om_{\bf k}^2) ,
\lab{calVdefbis}
\ee
which is the familiar phase space density factor of finite temperature
field theory.  

With a view to the interpretation of the $X_\veck$ operators, observe that
there are some
bilinear normal-ordered combinations of the $X_{{\bf k}}$ which are {\em
not} zero when thermal averaged; for example
\be \lab{z1}
\ll X^\flat_{{\bf k}}  X^\flat_{-{\bf k}} \gg 
= - \ll X_{{\bf k}}  X_{-{\bf k}} \gg = \sqrt{ n_k(1+n_k)} \; .
\label{XXexpect}
\ee
This represents a non-trivial problem in any attempt to 
interpret the
$X_{{\bf k}}$ operators in terms of some type of new thermal
excitation.  However, such analysis is better posed in the language of
bilinear transformations \cite{TFD,BK} including Bogoliubov
transformations.  In any case the split \tseref{ppr} is reminiscent of
structures seen with Bogoliubov transformations.  We now rewrite the
$X_\veck$ operators in this language.

\section{The canonical transformation structure}

{}From the definition {\bf }of $X_\veck$ and $X_\veck^\flat$, we see that they
 mix $a_\veck$ and $a_{-\veck}^\dagger$ operators carrying opposite
momentum.\tsenote{even though we still have a canonical set if in the
 $X_\veck$'s we mix $a$ and $a^\dagger$ operators with
the same momentum.  In that case the $X_\veck$'s pick up and ${\bf x}$
dependence and I think the transformation is now a normal squeezing
rather than Bog} Such a mixing appears often in many other
contexts, e.g.\ cosmological perturbations \cite{Pr}  and
BCS theory 
\cite{FW}.\tsenote{\cite{FW}, equation (35.12).}  
The important point is that for any given $\veck
\neq 0$, $a_{\bf k}, a_{\bf k}^{\dagger}$ commute with $  a_{-{\bf
k}}, a_{-{\bf k}}^{\dagger}$; more generally
\beq
[ a_{\veck} , a^\dagger_{\vecp} ] = \delta_{\veck , \vecp} , \; \;
\; [ a_{\veck} , a_{\vecp} ] = [ a^\dagger_{\veck} , a^\dagger_{\vecp} ] =
0  .
\eeq 
That is, for a given ${\bf k} \neq 0$, we think of the sets
$\{a_{\bf k}, a^{\dagger}_{\bf k}
\}$ and
$\{a_{-\bf k}, a^{\dagger}_{-\bf k} \}$ as being independent.  
Canonical transformations mixing the two sets \cite{Pr,FW} arise
naturally if we 
%Such a partition of
%the space of all $a_{{\bf k}}$ into two
%parts\footnote{One must exclude the case $\veck=0$ and we always assume
%that this is done in what follows.} 
%is a natural construction, as can be seen by 
consider the Fourier
amplitudes of the real field $\phi$ and its 
conjugate $\pi (x) = \dot \phi (x)$;
\be
\phi(x) = {\frac {1}{\sqrt{V}} } \sum_{ \bf k}
q_{\bf k}(t)e^{i {\bf k}.{\bf x}}
\lab{four1},
\ee
\be
\pi (x) = {\frac {1}{\sqrt{V}} } \sum_{ \bf k}
p_{\bf k}(t)e^{-i {\bf k}.{\bf x}}
\lab{four2}.
\ee 
Here
\be
q_{\bf k}(t) = {1\over{(2{\om}_{\bf k})^{1\over 2}}}\le( a_{\bf 
 k}(t) + a_{-{\bf k}}^{\dagger}(t) \ri), \; \; \; \; \; \;
p_{\bf k}(t) = i\le( {\om_{\bf k}\over 2} \ri)^{1\over 2} \le( 
a_{\bf k}^{\dagger} (t) -  a_{-{\bf k}}(t) \ri), 
\lab{pqpos}
\ee
\be
q^{\dagger}_{\bf k}(t) = q_{- \bf k}(t), \; \; \; \; \; \; 
p^{\dagger}_{\bf k}(t) = p_{- \bf k}(t)
\lab{pqneg},
\ee
and 
\be
[q_{\bf k}(t), p_{\bf k'}(t)] = i{\delta}_{{\bf k},{\bf k'}}
\lab{fourcomm}.
\ee
Observe first that by definition $p_{\bf k}$ and $q_{\bf k}$ mix
creation and annihilation operators with opposite
momenta.  Secondly, 
%
%Our aim is to use this partition of ${\bf k}$-space into two commuting
%sets and look for canonical transformations which mix the two halves.  Let us
%briefly review some well known \cite{BK} and some less well known
%possibilities.  Thinking in terms of the $q$ and
%$p$ operators and their commutation relations, 
%one possible canonical transformation involves
%
consider a {\em rescaling} operation of the $q$ and $p$
operators: 
$q_{\bf k} \rightarrow e^{-\theta_{\bf k}}q_{\bf k}$, 
$p_{\bf k} \rightarrow e^{\theta_{\bf k}}p_{\bf k}$.  
This preserves the commutation
relation \tseref{fourcomm} for {\em any} function
$\theta_{\veck}$.  However, in order to preserve the hermitian
relationship \tseref{pqneg} between $q$
and $p$ for positive and negative $\veck$, 
one must choose $\theta_{\veck}=\theta_{-\veck}$.
Then
the rescaling of $q$ and $p$ generates a Bogoliubov transformation amongst the
creation and annihilation operators;
\begin{eqnarray}
\ak \rightarrow & b_{\veck}(\theta) 
&= e^{-i G_B(\theta)} \ak e^{iG_B(\theta)} =
\ak \cosh \theta_{\bf k} - \admk \sinh \theta_{\bf k},
\label{aB1}
\\
\adk \rightarrow & b^\dagger_{\veck}(\theta) 
&= [b_{\veck}(\theta) ]^\dagger,
\label{adB1}
\\
\amk \rightarrow & b_{-\veck}(\theta) 
&= e^{-i G_B(\theta)} \amk e^{iG_B(\theta)} =
\amk \cosh \theta_{\bf k} - \adk \sinh \theta_{\bf k},
\label{amB1}
\\
\admk \rightarrow & b^\dagger_{-\veck}(\theta) 
&= [b_{-\veck}(\theta) ]^\dagger,
\label{amdB1}
\end{eqnarray}
where
\be
G_B(\theta) = \frac{i}{2} \sum_{{\bf k}}\theta_{{\bf k}}
\le[a^{\dagger}_{{\bf k}}a^{\dagger}_{-{\bf k}}- 
a_{-{\bf k}}a_{{\bf k}}\ri]
\label{GBdef2}
\ee
and $[b_{\veck}(\theta),b^\dagger_{\bf p}(\theta)] = \delta_{\bf k,p}$
with all other commutators zero.\footnote{The factor
of $1/2$ is needed because $\theta_{\bf k}$ and $\theta_{-\bf k}$ are
regarded as the {\em same} independent variable.  The point ${\bf
k}=0$ is omitted from the summation.}
This transformation defines a new vacuum for the $b$ operators
\be
|0(\theta)\ran \ran = e^{-iG_B(\theta)}|0\ran \; , \; \; \; 
b_{\veck} (\theta) |0(\theta)\ran \ran = 0
\label{vac2}
\ee
which is orthogonal to $\langle 0|$ in the infinite volume limit. 
This leads to relations such as
\be
\sinh^2 (\theta_{{\bf k}} )
= \langle \langle 0(\theta)| a_{{\bf k}}^{\dagger}a_{{\bf k}}
|0(\theta)\ran \ran 
= \langle 0| b_{{\bf k}}^{\dagger}(-\theta)b_{{\bf k}}(-\theta) |0\ran ,
\label{thetaset}
\ee
where we have used the hermitian property of this transformation to
write down the two equivalent forms.  In the context of thermal field 
theories, the $\alpha = 1/2$ formulation TFD
\cite{Raybook,He,TFD} uses a similar Bogoliubov
construction.  There $\theta_\veck$ is related to the 
particle number distribution functions though
\tseref{thetaset}.

It is interesting to note, however, that rescaling $q$ and $p$ with
an odd function of $\theta_{\veck}$ leads to a straight rescaling of
the annihilation and creation operators,
\be
\ak \rightarrow a'_{\bf k}(\theta) = e^{ \sum_{\veck}\theta_{\veck}
N_{\veck}}
\ak e^{ -\sum_{\veck}\theta_{\veck} N_{\veck}}= e^{-\theta_{\veck}}
\ak ,
\label{asc2}
\ee
\be 
\adk \rightarrow  a^{' \, \natural}_{\bf k}(\theta) = e^{
\sum_{\veck}\theta_{\veck} N_{\veck}}
\adk e^{ -\sum_{\veck}\theta_{\veck} N_{\veck}}= e^{\theta_{\veck}}
\adk ,
\label{adsc2}
\ee
where the sum is over all momenta, and $N_{\bf k} =
{a}^{\dagger}_{\bf k} {a}_{\bf k}$.
Note that unlike the case of the Bogoliubov transformation 
\tseref{GBdef2}, this rescaling transformation does {\em not} mix
operators with opposite momentum.  
Thus in fact \tseref{asc2} and \tseref{adsc2} hold for all
momenta and for 
any function $\theta_{\veck}$.  However, only for odd functions does it
correspond to a rescaling of $q$ and $p$.  The important point to note
is that unlike the Bogoliubov transformation, this
transformation is {\em not} hermitian, and as a result it is not
usually discussed.  It does though lead to a canonical set of
operators; 
$ [a'_{\bf k}(\theta),a^{' \, \natural}_{\bf p}(\theta)] 
= \delta_{\bf k,p}$ with all other commutators zero.  Here the ``natural
conjugation'', $\natural$, consists of hermitian conjugation and the
replacement $\theta \leftrightarrow -\theta$ \tseref{adsc2}.
Amusingly, if we put $\theta_\veck =
\omega_\veck t_4$, this rescaling is a Euclidean time translation by
$t_4$, and such rescalings play a key role in all path ordered approaches to
thermal field theory.  They also appear in TFD for $\alpha \neq 1/2$
formulations \cite{He}.   Thus rescaling transformations are in
fact extremely common.\tsenote{While this 
transformation may have lost its hermitian properties, it
does have the powerful property that it preserves the 
Weyl-Heisenberg algebra - the $ \{ a, a^\dagger, H = [a,
a^\dagger]/2, 
N =  a^\dagger a \} $ and its commutation relations - provided one
replaces $\dagger$ with $\natural$.  This is
obvious since the rescaling 
transformation takes the form of 
$T.(\mbox{operator}).T^{-1}$ for all
operators, even though $T^{-1} \neq T^\dagger$.}   

Having turned away from hermitian transformations,
we can return to the Bogoliubov transformation
\tseref{aB1}-\tseref{amdB1} and note
that there is another closely related canonical pair related by the
$\natural$ operation:
\bea
\ak \rightarrow 
&  c_{\veck}(\theta) & =  \frac{1}{m_{\bf k}} b_{\veck}(\theta) 
= \frac{1}{m_{\bf k}}e^{-i G_B(\theta)} \ak e^{iG_B(\theta)} 
\nonumber \\ 
&& 
=\frac{1}{m_{\bf k}}
\left[ \ak \cosh \theta_{\bf k} - \admk \sinh  \theta_{\bf k} \right],
\label{aB2}
\\
\admk \rightarrow 
&  c^\natural_{-\veck}(\theta) & = \frac{1}{m_{\bf k}} [b_{-\veck}(-\theta)]^\dagger
= \frac{1}{m_{\bf k}}
e^{i G_B(\theta)} \admk e^{-iG_B(\theta)} 
\nonumber \\ 
&& 
 = \frac{1}{m_{\bf k}} \left[ \ak \sinh \theta_{\bf k} + \admk \cosh 
 \theta_{\bf k}
\right] ,
\label{amdB2}
\\
\amk \rightarrow 
& c_{-\veck}(\theta) & = \frac{1}{m_{\bf k}} b_{-\veck}(\theta)
= \frac{1}{m_{\bf k}} e^{-i G_B(\theta)} \amk e^{iG_B(\theta)} 
\nonumber \\ 
&& 
 =\frac{1}{m_{\bf k}} \left[ 
\amk \cosh \theta_{\bf k}  - \adk \sinh  \theta_{\bf k} \right],
\label{amB2}
\\
\adk \rightarrow 
& c^\natural_{\veck} & =  \frac{1}{m_{\bf k}} (b_{\veck}(-\theta))^\dagger
= \frac{1}{m_{\bf k}} e^{i G_B(\theta)} \adk e^{-iG_B(\theta)} \nonumber \\ && 
= \frac{1}{m_{\bf k}} \left[ \amk \sinh \theta_{\bf k}  + \adk \cosh  \theta_{\bf k}
\right] ,
\label{adB2}
\eea
with
\be
 m_\veck = m_{-\veck}= 
\left[\sinh^2 \theta_{\bf k}  + \cosh^2 \theta_{\bf k} 
\right]^{1/2} ;
\ee
where we have used $G_B(-\theta) = - G_B(\theta)$.  Observe that the
$\natural$ conjugation is defined as above, and that it is
responsible for the
change of sign in front of $\sinh \theta$ terms in moving
from \tseref{aB2} to \tseref{amdB2}, or from 
\tseref{amB2} to \tseref{adB2}.  It is also responsible for the origin
of the 
normalisation, $m_{\bf k}$. 
This transformation preserves all the commutation relations
\tsenote{despite the fact that in this case a straight substitution of the
$\exp \{ \pm i G_B \}$ transformations into the $a$ commutators does not
appear to leave them invariant.  Indeed unlike the rescaling
transformation, this Bogoliubov-like transformation is not even of
the form $T.(\mbox{operator}).T^{-1}$. What happens to the
Weyl-Heisenberg algebra and Hopf algebra structure?}
\beq
[ c_{\veck} , c^\natural_{\vecp} ] = \delta_{\veck , \vecp} , \; \;
\; [ c_{\veck} , c_{\vecp} ] = [ c^\natural_{\veck} , c^\natural_{\vecp} ] =
0.
\eeq

Now we return to the case of the $X_{\bf k}$ operators \tseref{X}.  The analysis 
is made much simpler by introducing
a specific temperature dependent angle $\si_{\veck} = \si_{-\veck}$
to parameterise the $T>0$ split defined by \tseref{fgdef1}.  Note that
$f_{\bf k}$ and $g_{\bf k}$
involve the Bose-Einstein distributions 
through factors like $n_{\bf k}$, $1+n_{\bf k}$ and their square
roots.   In many situations in thermal field theory the special properties of these
distributions are crucial and may be encoded by the use of
hyperbolic functions.  Thus we are led to write
\be
n_{{\bf k}} = \sinh^2 ( \si_{{\bf k}} ) = \frac{1}{e^{\bt \om_{\bf k}}-1},
\ee
though this does not fix the sign of $\sigma_k$.
Again, TFD is especially inspirational as it uses the same
parameterisation for the Bose-Einstein distribution 
$n_{\bf k}$ \cite{Raybook,LvW,vW,He,TFD}
 though it applies it in a different way.

In terms of $\si_{\veck}$, the functions 
$f_{\bf k}$ and $g_{\bf k}$ of \tseref{fgdef1} 
can be rewritten as\tsenote{The role of the sign $s_\veck$ can be
completely taken over by the arbitrary nature of the sign of
$\sigma_\veck = | \sigma_\veck | s_\veck$.}
\bea
f_{\bf k}= e^{-\si_{\bf k}}\;\sinh(\si_{\bf k}), && 
g_{\bf k}= -e^{\si_{\bf k}}\;\sinh(\si_{\bf k}), 
\label{fgdef}
\\
(1-f_{\bf k})= e^{-\si_{\bf k}}\;\cosh(\si_{\bf k}),
&&
(1-g_{\bf k}) = e^{\si_{\bf k}}\;\cosh(\si_{\bf k}).
\eea
This shows that changing the sign of $\si_{\bf k}$ is equivalent to
swapping the $f_{\bf k}$ and $g_{\bf k}$ functions.  Thus 
$\flat$-conjugation is hermitian  conjugation plus the exchange
${\si}_{\bf k} \lrar -{\si}_{\bf k}$; that is, for some operator
$A[\sigma]$ we have
\beq
\left( A[\sigma_\veck] \right)^\flat = 
\left( A[-\sigma_\veck] \right)^\dagger .
\label{flatdef2}
\eeq
Thus we see that the $\flat$ and $\natural$ operation are identical
if we set $\theta = \sigma$ in the definition of $\natural$.

We may now re-write the $X_\veck$ operators of \tseref{X}
in terms of $\sigma_\veck$ (dropping the $t$-dependence for notational
simplicity);
\be
 X_{\bf k} = \frac{1}{d_{\bf
k}}\left(\cosh(\si_{\bf k}) e^{- \si_{\bf k}}\ak - \sinh(\si_{\bf k})
e^{\si_{\bf k}} \admk \right),
\label{Xexplicit}
\ee
\be
 X^{\flat}_{\bf k} = \frac{1}{d_{\bf k}}
\left( \sinh(\si_{\bf k}) e^{-\si_{\bf k}} \amk 
+ \cosh(\si_{\bf k}) e^{\si_{\bf k}} \adk \right),
\label{Xfexplicit}
\ee
where $\sigma_\veck = +\sigma_{-\veck}$, $s_\veck = +s_{-\veck}$. 
From these equations it is straight forward to see that we can now view the $X_{\bf k}$ operator
as being generated by the combination of a scaling transformation, 
which generates the parts $e^{-\si_{\bf k}}\ak$ and $e^{\si_{\bf k}}
\admk$ of \tseref{Xexplicit} according to \tseref{asc2} and \tseref{adsc2},
and the new transformation of \tseref{aB2}-\tseref{adB2}.  
Thus, depending on the order in which we carry out these two
operations, we may write
\begin{eqnarray}
 X_{\bf k} & = & \frac{1}{d_{\bf k}}
\left( 
e^{-i G'_B(\si)} \;
 e^{ \sum_{\veck}\si_{\veck} N_{\veck}} \;
\ak \;
e^{ -\sum_{\veck}\si_{\veck} N_{\veck}} \;
e^{iG'_B(\si)}
 \right)
\label{Xexplicitfi1}
\\
& = & \frac{1}{d_{\bf k}}
\left( 
 e^{ \sum_{\veck}\si_{\veck} N_{\veck}} \;
e^{-i G_B(\si)} \;
\ak \;
e^{i G_B(\si)} \;
e^{ -\sum_{\veck}\si_{\veck} N_{\veck}} 
\right)
\label{Xexplicitfi2}
\end{eqnarray}
and
\begin{eqnarray}
 X^{\flat}_{\bf k} & = & \frac{1}{d_{\bf k}}
\left( 
e^{i G'_B(\si)} \;
 e^{ \sum_{\veck}\si_{\veck} N_{\veck}} \;
\adk \;
e^{ -\sum_{\veck}\si_{\veck} N_{\veck}} \;
e^{-iG'_B(\si)} 
 \right)
\label{Xexplicitffi1}
\\
& = & \frac{1}{d_{\bf k}}
\left( 
 e^{ \sum_{\veck}\si_{\veck} N_{\veck}} \;
e^{i G_B(\si)} \;
\adk
e^{-i G_B(\si)} \;
e^{ -\sum_{\veck}\si_{\veck} N_{\veck}}
\right).
\label{Xexplicitffi2}
\end{eqnarray}
with 
\beq
d_\veck = d_{-\veck} = (\cosh (2 \sigma_\veck) )^{1/2} = \left( 1 + 2
n_{\bf k} \right)^{1/2}.
\eeq
Here 
\be
G'_B(\si)=\frac{i}{2}\sum_{{\bf
k}}\si_{{\bf k}} 
\le[{a'}^{\flat}_{{\bf k}}
{a}'^{\flat}_{{\bf -k}}- {a'}_{-{\bf k}}{a'}_{{\bf k}}\ri]
=-(G_B^{'\flat}(\si))^{\flat},
\ee
 where $a^{'}_{\bf k}$ and 
$a^{' \, \flat}_{\bf k}$ are
implicitly functions of $\sigma$ and are defined in \tseref{asc2} and
\tseref{adsc2} (with $\theta \rightarrow \sigma, \natural
\rightarrow \flat$).  Thus finally we see that
the new $X_\veck$'s are related to the original $a_\veck$'s by a
combination of a Bogoliubov-like transformation \tseref{aB2}-\tseref{adB2} and a scaling transformation
\tseref{asc2}-\tseref{adsc2}. 

We can also construct the vacuum, $| 0_{X}(t) \rangle$, 
for the $X_\veck$ operators
\bea
X_{\bf k}(t) |  0_{X}(t) \rangle &=& 
\langle   0_{X}(t) | X^\flat_{\bf k}(t) = 0 .
\eea
From \tseref{Xexplicitfi1} we see that the ket vacuum must be
\bea
| 0_{X}(t) \rangle & = & N e^{-iG'_B(\sigma)}|0\ran 
\\
& = &N \prod_{{\bf k}}
\frac{1}{\cosh (\si_{{\bf k}} )} 
\exp\le[\tanh ( \si_{{\bf k}} ) {a'}_{{\bf k}}^{\flat}(t)
{a'}_{-{\bf k}}^{\flat}(t)\ri]|0\ran  .
\lab{w6}
\eea
The $X$ vacuum is therefore a condensate of zero-momentum pairs of
$a$-particles\tsenote{an 
$su(1,1)$ (time dependent) generalised coherent state?} 
In defining the
bra vacuum, care must be taken to ensure $\flat$-
conjugation is consistently used rather than hermitian conjugation.
Thus from \tseref{Xexplicitffi1} we can define
\bea
\langle  0_{X}(t) | &= &(|  0_{X}(t)\rangle)^{\flat} \; 
\neq \; (|  0_{X}(t) \rangle )^\dagger \; = \;  \lan 0 | e^{+iG'_B(\sigma)} N
\\
&=&
\lan 0|e^{-iG_B'(\sigma)} N
\nonumber
\\
& =& \lan0| N \prod_{{\bf k}}
\frac{1}{\cosh (\si_{{\bf k}} ) } 
\exp \le[ - \tanh ( \si_{{\bf k}} ) {a'}_{{\bf 
k}}(t) {a'}_{-{\bf k}}(t)\ri] .
\lab{w7}
\eea
The normalisation factor $N$ is not trivial and we find\tsenote{M.
thinks it should be $d_{\veck}^{2}$ and I think it may be
$d_{\veck}^{-1}$.}
\bea
\langle   0_{X}(t)| 0_{X}(t) \rangle  & = & 1 
\; \; \; \Rightarrow \; \; \;
N =  \prod_{\bf k} d_{\veck}^{2}  \; .
\eea
In the infinite volume limit we have
\bea
\langle  0_{X}(t) | 0 \rangle \; , \; 
\langle 0 | 0_{X}(t) \rangle  &\rar& 0 \; \; \; 
\forall \; t, \; \; V \rightarrow \infty.
\label{overlap}
\eea

We now have the full structure of the new operators $X_\veck$ and their
associated Fock space.  At this stage in the usual examples of
Bogoliubov transformations (symmetry breaking, TFD, etc.) we would
show that the physical vacuum was the transformed vacuum and
not that associated with the physical operators $a$. 
When we attempt to do this here, we find that
for a general operator $A[a,a^\dagger ]$ we have
\beq
\, \langle 0_{X}(t) |A[a,a^\dagger ]|  0_{X}(t) \rangle \neq \ll {A} \gg.
\eeq
For example
\bea
 \frac{1}{\langle   0_{X}(t) |  0_{X}(t) \rangle }
\langle   0_{X}(t) |a^\dagger_\veck  a_\veck |  0_{X}(t) \rangle 
&=& \frac{-1}{\exp ( \beta \omega_k) +1}
%- \frac{n_\veck}{d_\veck)}  
\neq n_\veck
\\
 \frac{1}{\langle   0_{X}(t) |  0_{X}(t) \rangle }
\langle   0_{X}(t) |a_\veck a^\dagger_\veck  |  0_{X}(t) \rangle 
&=&  \frac{1}{\exp ( - \beta \omega_k) +1}
%\frac{(1+ n_\veck)}{d_\veck)}
\neq 1+ n_\veck
\eea
Hence the vector $| 0_X(t) \ran$ (and
$\,\lan  0_X(t) |$) {\em is not a thermal vacuum}.\tsenote{But if we
looked at the expectation value of the {\em non-canonical} 
operators $a_\veck/d_\veck$
then we would almost have the right structure!}

\section{Conclusion and Outlook}

In this paper we have discovered a new set of
canonical operators for real scalar fields in thermal equilibrium --
the $X_\veck$'s of \tseref{Xexplicit} and \tseref{Xfexplicit} -- for all
path ordered approaches to thermal field theory such as Matsubara's
imaginary time-formalism.  These follow from the redefinition of the
normal ordered product found necessary if the canonical approach
path-ordered thermal field theory is to proceed in the usual way
\cite{ES}.  We have then shown that the $X_\veck$'s are produced from
the original creation and annihilation operators by a pair
of transformations, one Bogoliubov-like and one rescaling
\tseref{Xexplicitfi1} and \tseref{Xexplicitffi2}.  Finding these
transformations is not trivial since the conjugate pair $X_\veck$ and
$X_\veck^\flat$ are not related by hermitian conjugation but by our flat
conjugation, \tseref{flatdef2}, so we have had to
look beyond the standard transformations of the literature. 

Through the use of Bogoliubov transformations in QFT, one is usually
able to talk in terms of quasi-particles, or in the case of TFD one
is able to replace the thermal trace with a thermal vacuum. 
Unfortunately, both the lack of a hermitian structure and the
inability to duplicate the thermal trace results using the vacuum of
the $X_\veck$ operators means that  it is difficult to give a
meaning in terms of physical excitations   to this new set of
canonical operators $X_{\bf k}$. This is not   surprising in view of
the mentioned difficulties of defining the particle   concept at
finite temperature. What is relevant, and in some sense it is   our
main result, is that nevertheless, the formalism may support a  
canonical transformation structure.  Hence, as we have shown,   well
defined vacuum state (and associated Hilbert space) exists. 
However, such a vacuum is not the thermal   vacuum. At same time,
the split in terms of the $X_{\bf k}$, equations (\ref{phiplus}),
(\ref{phiminus}) and (\ref{ppr}), guarantees that the thermal
averages of all $n$-point normal ordered products of fields
vanish, which in turn makes the path-ordered approach to thermal
field theory to proceed in the usual way \cite{ES}.

Finally, there are some other results which are worth noting.  There
are in fact several alternative sets of canonical operators based on
the thermal normal ordered product.  One could just mix $a_\veck$
and $a^\dagger_\veck$ but keep the same coefficients rather than mix
$a_\veck$ and $a^\dagger_{-\veck}$ as we considered in \tseref{X}. 
This, however, would make the $X_\veck$'s contain a specific
position dependence, $X(x)$.  Alternatively, one can perform a
further rescaling of the $X_\veck$'s given here, $X_\veck
\rightarrow \exp \{ \theta_\veck \} X_\veck $ , $X^\flat_\veck
\rightarrow \exp \{ -\theta_\veck \} X^\flat_\veck $ for any
$\theta_\veck$  and keep the commutation relations.  

Most interestingly one can exploit the freedom in the sign of
$\sigma_\veck$ and work with an odd function rather than the even
function we have chosen here (equivalent to $s_\veck= - s_{-\veck}$
in \tseref{fgdef1}).  Still using the idea that annihilation and
creation operators of opposite momenta form mutually commuting sets,
we mix just the annihilation operators (or just creation operators)
of opposite momentum to form new canonical operators, $W$.  
This technique works for all
types of bosonic field.  For example,  in the case of
non-relativistic fields we can define four new canonical operators $\Wp$,
$\WFm$, $\Wm$ and $\WFp$ based on the split given in \cite{ES}.
There is some flexibility in the notation one can use.  If we
stick with the flat conjugation definition \tseref{flatdef2} and
demand that normal ordering of \tseref{N2} and \cite{ES} 
puts the annihilation operators
$W_1,W_2$ to the right and creation operators $W^\flat_1,W^\flat_2$
to the left, then we find that
\begin{eqnarray}
{\psi}^{(+)}(x) =&
\sum_{\veck} \fnorm
\left[ (1 - f_{\veck}) {a}_{\veck} e^{-i\omega t}e^{i\veck.\vecx} 
\right] &= 
\sum_{{\bf k}>0} 
\frac{\cosh ( \sigma_\veck ) }{(\omega_{\bf k} V)^{\half} }
\Wp (\veck,x)e^{-i\omega t}
\\
{\psi}^{(-)}(x) =& 
\sum_{\veck} \fnorm
\left[ f_{\veck} \A_{\veck} e^{-i\omega_{\bf k} t}e^{i\veck.\vecx}  
\right]
& =\sum_{{\bf k}>0} 
\frac{\sinh ( \sigma_\veck ) }{(\omega_{\bf k} V)^{\half} }
\WFm (\veck,x)
e^{-i\omega_{\bf k} t}
\\ {\psi}^{\dagger(+)}(x) =& \sum_{\veck} \fnorm \left[  
g_{\veck} {a}^{\dagger}_{\veck} e^{-i\omega_{\bf k} t}e^{i\veck.\vecx} 
\right] 
&= \sum_{{\bf k}>0} 
\frac{\sinh ( \sigma_\veck ) }{(\omega_{\bf k} V)^{\half} }
\left[ -\Wm (\veck,x) \right] e^{-i\omega_{\bf k} t}
\\ {\psi}^{\dagger(-)}(x) =& \sum_{\veck} \fnorm
\left[  (1-g_{\veck}) {a}^{\dagger}_{\veck} e^{-i\omega_{\bf k} t} 
e^{i\veck.\vecx} \right] 
&= \sum_{{\bf k}>0}
\frac{\cosh ( \sigma_\veck ) }{(\omega_{\bf k} V)^{\half} }
 \WFp (\veck,x)e^{-i\omega_{\bf k} t}.
\end{eqnarray}
where the sum over $\veck >0$ indicates that we are summing over
half of $\veck$ space, including only one of each $(\veck,-\veck)$
pair. 
The commutation relations satisfied by the $W$'s are then seen to be
\begin{equation}
\left[ W_{i}(\veck,\vecx), W_{j}^{\flat}(\vecp,\vecx) \right] 
= \left(
\begin{array}{cc} 1 & 0 \\ 0 & -1 \end{array} \right)  \delta_{\veck,\vecp} 
\label{Wcommrel}
\end{equation}
with other commutators zero.  We are currently investigating the
structure defined by these operators and its relation to that defined
by the $X_\veck$'s.

\section*{Acknowledgements}

T.S.E. thanks  the
Royal Society for their support.  D.A.S. is supported by P.P.A.R.C.\
of the U.K.  This work was supported in part
by the European Commission under the Human Capital and Mobility
programme, contract number CHRX-CT94-0423.

% **********************************************************************


\begin{thebibliography}{99}

\bibitem{ES} T.S.Evans and D.A.Steer,  Nucl.Phys. {\bf B474} (1996) 481
%Imperial/TP/95-96/18
\eprintpub{hep-ph/9601268}. 

\bibitem{Ma} T.Matsubara, {Prog.\ Th.\ Phys.\ }  {\bf 14} (1955) 4.

\bibitem{Raybook} {R.J.Rivers}, Path Integral Methods in Quantum
Field Theory (Cambridge University Press, Cambridge, 1987).

\bibitem{LvW} N.P.Landsman and Ch.G.van Weert,  Phys.\ Rep.\ {\bf 145}
(1987) 141.

\bibitem{vW} Ch.G. van Weert, in: Proceedings of the Banff/CAP
Workshop on Thermal Field Theories ed.\ F.C.Khanna, R.Kobes,
G.Kunstatter, H.Umezawa (World Scientific, Singapore, 1994) p.1.

\bibitem{LeB} M. Le Bellac, Thermal Field Theory
(Cambridge University Press, Cambridge, 1996).

\bibitem{St} F.Strocchi, Elements of Quantum Mechanics of Infinite
Systems (World Scientific, Singapore, 1985).

\bibitem{mix} M.Blasone, P.A.Henning and G.Vitiello, ``Mixing
Transformations in Quantum Field Theory and Neutrino Oscillations'',
in Proceedings of "Results and Perspectives in Particle
Physics", La Thuile, Aosta Valley, March 1996, M.Greco Ed., INFN Pub. 
Frascati 1997 \eprint{hep-ph/9605335} ; 
E. Alfinito, M. Blasone, A. Iorio and G. Vitiello, 
Acta Phys.Polon.{B27} (1996) 1493 \eprintpub{hep-ph/9601354};
M.Blasone and G.Vitiello, {Annals of Physics}  {244} (1995) 283.

\bibitem{MSV} M.Martelini, P.Sodano and G.Vitiello,
Nuovo Cimento {\bf A48} (1978) 341.

\bibitem{BD}  N.D.Birrell and P.C.W.Davis, Quantum Field in Curved Space
Time (Cambridge University Press, Cambridge, 1988).

\bibitem{AMV} E.Alfinito, R.Manka and G.Vitiello, ``Double Universe''
\eprint{hep-th/9705134}.

\bibitem{QD} E.Celeghini, M.Rasetti and G.Vitiello,
Ann.\ Phys.\ {\bf 215} (1992) 156;
 A.Iorio, G.Vitiello, Ann.\ Phys.\ {\bf 241} (1995) 496 \eprintpub{hep-
th/9503136}; Y.N. Srivastava, G. Vitiello and A. Widom, 
Ann.\ Phys.\ {\bf 238} (1995) 200 \eprintpub{hep-th/9502044}.

\bibitem{BV} N.N.Bogoliubov, Sov.Phys.-JETP {\bf 7} (1958) 41; 
J.G.Valatin, Nuovo Cimento {\bf 7} (1958) 843.

\bibitem{ja} R.Jackiw, Two lectures on two-dimensional gravity 
\eprint{gr-qc/9511048}; D.Cangemi, R.Jackiw and B.Zwiebach, Annals of
Physics (N.Y.) {\bf 245} (1996), 408. 

\bibitem{La} N.P.Landsman, Ann.Phys. {\bf 186} (1988) 141.

\bibitem{He} P.A.Henning, Phys.\ Rep.\ {\bf 253} (1995) 235.

\bibitem{TFD} Y.Takahshi and H.Umezawa, Collective Phenomena {\bf 2} 
(1975) 55;
H.Umezawa, H.Matsumoto and M.Tachiki, Thermo Field
Dynamics and Condensed States, (North Holland, Amsterdam, 1982).

\bibitem{Louis} M.Schmutz, Z.\ Phys.\ {B30} (1978) 97; 
I.D.Lawrie, J. Phys. A27 (1994) 1435. 

\bibitem{BK} {S.Barnett and P.Knight}, J.\ Opt.\ Soc.\ A, {\bf B21} (1985)
467; B.L.Schumaker, Phys.\ Rep.\ {\bf 135} (1986) 317.

\bibitem{Pr} T.Prokopec, Class.\ Quantum Grav.\ {\bf 10} (1993) 2295.

\bibitem{FW} {A.Fetter and J.Walecka}, Quantum Theory of
Many-Particle Systems (McGraw-Hill, New York, 1971).

\end{thebibliography}
\end{document}